\documentstyle[preprint,aps,version2]{revtex}
\newcommand{\beq}{\begin{equation}}
\newcommand{\eeq}{\end{equation}}
\newcommand{\beqar}{\begin{equation}\begin{array}}
\newcommand{\eeqar}{\end{array}\end{equation}}

\def\rd{{\rm  d}}

\def\bp{{\bf  p}}
\def\rtr{{\rm Tr}}

\begin{document}
\draft

\vspace{30mm}

\begin{center}
{\large
{
The strange quark spin of  the proton
}

{
in  semi-inclusive $\Lambda$  leptoproduction
\footnote
{This work is supported in part by the Chinese National Science
Foundation under grant No.~19445004.}
}
}
\end{center}

\vspace{10mm}

\author{Wei Lu and Bo-Qiang Ma}

\begin{instit}
CCAST (World Laboratory), P.O. Box 8730, Beijing 100080, China

		and

Institute of High Energy Physics, Academia Sinica,
P.O. Box 918(4), Beijing 100039, China\footnote{Mailing address}
\end{instit}

\begin{abstract}

Considering the possible interpretation of the
Ellis-Jaffe sum rule violation
that the strange quark in the  polarized
proton is remarkably polarized in the opposite
direction of the proton spin,
we investigate its implication in the semi-inclusive
$\Lambda$ production in deep
inelastic process
with the  electron beam unpolarized and proton
target polarized longitudinally.  As a result, we find
that the measurement of the $\Lambda$ polarization
in the process considered
can provide clean information
about the strange quark spin distribution in the proton.
\end{abstract}

\centerline{{\it to be published in Phys. Lett. B}}
\newpage

Since the discovery of the Ellis-Jaffe sum rule violation
by the European Muon Collaboration (EMC) experiment \cite{EMC88},
the proton  spin structure has received extensive attention.
The small EMC data of the integrated spin structure function,
in combination with the Bjorken sum rule, is generally understood
to imply that the sum of the up, down, and strange
 quark helicities  in the proton
is much smaller than the proton spin \cite{Spin}.
During the past years, there have appeared
many theoretical works on the proton spin content and
the Ellis-Jaffe sum rule violation \cite{review}.
One of the possible interpretations is
that the strange quark
content in a longitudinally polarized proton may be
polarized antiparallel to the proton
polarization direction \cite{Strange}.
It is clear that further experimental information
about  the strange quark helicity distribution in the
proton will  reveal  more about the proton spin structure
and test  the various interpretations of the Ellis-Jaffe
sum rule violation.
Recently, Alberg, Ellis, and Kharzeev \cite{AEK}  have proposed that
the measurement of target spin depolarization parameter in the
$\bar p p \to \bar \Lambda \Lambda$ construct a test of the
polarization state of the strange quarks in the proton.

The semi-inclusive processes in the deep inelastic
scattering seem to be a clean process for providing further
information about the distributions of the
different  quark flavors  in the proton \cite{SIP}.
In this Letter, we investigate the implication of the
possible large negatively-polarized strange sea content
in the proton to the semi-inclusive  $\Lambda$ production
in the current fragmentation region, with the lepton
beam unpolarized and the proton target polarized longitudinally.
To be specific, we consider
\begin{equation}
l +p (P,S) \to l^\prime
+\Lambda(P_\Lambda, S_\Lambda) +X,
\end{equation}
with the proton target polarized along  the direction of
the lepton beam direction. We normalize the spin vector
in such a way that $S\cdot S=-1$ for a pure spin-half fermion state.

Our motivation is as follows. If the strange quark $is$
negatively polarized in a polarized proton,
such a polarization should be reflected in the final
state, although such information is regrettably  summed over
in the sense of the deep inelastic scattering.
Considering that the quark
 helicity is conserved in the
underlying high-energy partonic subprocesses,
the strange quark polarization,
if any,  should remain unchanged after its deeply scattering by the
incident virtual photon.
Then, the polarization of the scattered strange quark
can  be inherited, at least partially,  by the
inclusively detected spin-half baryons in  its fragmentation products.
In principle, the strange quark energized by the
virtual photon will
fragment into a strange hadron plus something else. Among the
produced strange particles,
we choose to study the $\Lambda$ hyperon simply because
it has the largest production cross section among hyperons.
In addition, the $\Lambda$ polarization is self-analysing
owing to its characteristic decay mode $\Lambda \to p \pi^-$
with a large branching ratio of 64\% \cite{booklet}.
  If the strange quark
is  indeed  negatively  polarized, one can  very intuitively
expect  the inclusive detected $\Lambda$  hyperons to
epitomize the polarization of  the parent strange quark,
no matter how complicated the  quark fragmentation process is.
In principle,  the polarization  of the scattered
 strange quark can also be self-analysed by other hyperons,
but the corresponding statistics will be poor since  they
have to be reconstructed from the $\Lambda$ and other particles.

In the one-photon  exchange approximation,  the  proton structure is
probed by a space-like photon with momentum $q=l-l^\prime$.
The invariant cross section for the process considered can be
written as a contraction of  the leptonic and hadronic tensors:
\beq
E^\prime  E_\Lambda \frac{\rd\sigma (S,S_\Lambda) }
{\rd^3 l^\prime  \rd^3 p_\Lambda}
= \frac{\alpha^2}{16 \pi^3 MEQ^4} L_{\mu\nu}(l,
l^\prime) W_{\mu\nu}(q,P,S,P_\Lambda,S_\Lambda),
\label{cross}
\eeq
where  $M$ is the target mass,
$E$ the beam energy in the laboratory frame,
and $Q=\sqrt{-q^2}$.

Conventionally, the leptonic and hadronic tensors are
defined as
\beq
L_{\mu\nu}(l,l^\prime )=\frac{1}{2} \rtr [\rlap/l^\prime \gamma_\mu
\rlap/l \gamma_\nu]
= 2l_\mu l^\prime_\nu
+2l^\prime_\mu l_\nu -Q^2 g_{\mu\nu},
\eeq
and
\begin{eqnarray}
W_{\mu\nu}(q,P,S,P_\Lambda,S_\Lambda)&=&
\frac{1}{4\pi}
\sum \limits_X
\int \rd^4 y \exp(i q \cdot y)
\nonumber \\
& & \times
<P,S|j_\mu (0)|\Lambda(P_\Lambda, S_\Lambda),X>
<\Lambda(P_\Lambda, S_\Lambda),X|j_\nu(y)|P,S>,
\end{eqnarray}
respectively.  The  electromagnetic current is defined  to
be $j_\mu=\sum\limits_f q_f \bar\psi_f\gamma_\mu\psi_f$
with $q_f$ being the quark charge in unit of the electron charge
and  $f$ being the flavor index.
According to the valence quark composition of the $\Lambda$
hyperon, we simply assume it is predominantly generated
in the strange quark fragmentation jet. In other words,
the flavor index will be simply understood to be strange in our
discussion.

Experimentally, people usually define the  following  kinematical
scalar variables:
\beq
x_B=\frac{-q^2}{2 P\cdot q} ,~ y=\frac{P\cdot q}{P\cdot l},~
z=\frac{P\cdot P_\Lambda}{P\cdot q}.
\eeq
Correspondingly, the cross section for semi-inclusive process can
be  written as
\beq
\frac{\rd \sigma (S,S_\Lambda) }{\rd x_B \rd y  \rd z
\rd^2 \bp_{\Lambda\perp}}
=
\frac{ \alpha^2 y  }{8 \pi^2 z Q^2} L_{\mu\nu}(l,l^\prime) W^{\mu\nu}
(q, P, S, P_\Lambda, S_\Lambda).
\eeq

All the information about the nucleon structure and $\Lambda$ hyperon
production  is encoded in the hadronic tensor. Because
of color confinement, its perturbative approach is heavily
dependent on the  factorization of amplitudes (equivalently cross
section) into the long-distance parton matrix elements
and short-distance  partonic interaction coefficients
\cite{fact}. The
prediction power of QCD factorization  theorem  consists
in that  those nonperturbative matrix elements are universal
in various processes whereas the  short-distance coefficients
can be calculated perturbatively. As a consequence,  one can
use the data on matrix elements  from  a complete set
of reference processes to make predictions for other processes.
In our case of the semi-inclusive $\Lambda$ production,
the strange quark distribution in
the proton does not depend on whether we detect a $\Lambda$
hyperon in the target fragmentation region or not.
Moreover, the strange quark helicity distribution
is guaranteed  by the
factorization theorem to be the
same as that in the deeply inelastic scattering.

In essence, the factorization of  QCD high-energy amplitudes
is an expansion in terms of  a hard partonic scale.
At the leading power (twist),
the factorization approach corresponds to the
quark-parton model prescription, with the long-distance   parton
matrix   elements  interpreted as the  parton distribution
function in the hadron or  parton  fragmentation function
for  the hadron production.    In the process we consider,
the polarization of the
semi-inclusively detected $\Lambda$'s is a leading-twist spin asymmetry.
So we will  simply adopt the  quark-parton model prescription
 in order to protrude the main physics.

Let us define  $s^{h_s}_{h_p} (\xi)$  to be  the probability to find a
strange quark with momentum fraction $\xi$ and  helicity $h_s$ in  the
proton with helicity $h_p$.  For a longitudinally polarized proton
two  independent quark distribution functions can be taken as
\beq
s(\xi)= s^\uparrow_\uparrow(\xi)  +s^\downarrow_\uparrow (\xi),
\eeq
\beq
\Delta s(\xi)= s^\uparrow_\uparrow(\xi)  -s^\downarrow_\uparrow (\xi) ,
\eeq
where $\uparrow$ and $\downarrow$ represent the positive
and negative polarization along the corresponding
particle momentum direction.  Conceptually,
$s(\xi)$ and $\Delta s(\xi)$ can be interpreted as the
color-summed strange quark
 density and helicity in the proton, respectively,

 Now we perform a similar discussion of  the strange
quark fragmenting into the $\Lambda$ particle. Let
$D^{ h_\Lambda}_{ h_s} (\xi)$ denote the probability
to find a $\Lambda$ hyperon with helicity $h_\Lambda$ in the
decay products of a strange quark with helicity $h_s$, carrying
a fraction $\xi$  of the momentum of the parent strange quark.
Parity conservation  tells us
\beq
D^\uparrow_\uparrow  (\xi )
=D^\downarrow_\downarrow  (\xi ),
{}~
D^\uparrow_\downarrow  (\xi )
=D^\downarrow_\uparrow  (\xi ).
\eeq
Hence, there are only two independent   $s\to \Lambda$ fragmentation
functions, which  we chosen as
\beq
D(\xi)=
D^\uparrow_\uparrow  (\xi )
+D^\downarrow_\uparrow  (\xi ),
\eeq
\beq
\Delta D(\xi)=
D^\uparrow_\uparrow  (\xi )
-D^\downarrow_\uparrow  (\xi )
\eeq
Just like in the case of  parton distribution,
$D (\xi)$  is interpreted as the  spin-independent probability
to find a $\Lambda$ hyperon with momentum
fraction $\xi$ in the parent strange quark,
while $\Delta D (\xi )$  can be thought of as the
corresponding helicity distribution of the $\Lambda$ hyperon.

Utilizing the parton distribution  and fragmentation functions
introduced above,  one can easily  obtain
\beq
\rd N_{\Lambda}^{h_{\Lambda}}(x_B, z)\propto e^2_s
[s(x_B)~ D(z) + h_p h_\Lambda \Delta s (x_B) ~ \Delta  D(z)]
 \rd x_B \rd z
\label{res1}
\eeq
for the number of  the $\Lambda$' particles  with helicity $h_\Lambda$,
produced in the region of $x_B \to x_B +\rd x_B$  and $z\to z +\rd z$.
In obtaining Eq. (\ref{res1}),  the integration
over the scaled variable $y$ and the $\Lambda$ transverse
momentum  $\bp_{\Lambda \perp}$  with respect to the  virtual
photon momentum direction have  been performed.
Therefore, we have the following quark-parton
model expression for the  longitudinal polarization of the
semi-inclusive  $\Lambda$ particles:
\beq
P_\Lambda (x_B, z)=\frac{N_{\Lambda}^\uparrow (x_B, z)-
N_{\Lambda}^\downarrow (x_B, z)}
{ N_{\Lambda}^\uparrow (x_B, z)+ N_{\Lambda}^\downarrow (x_B, z)}
=\frac{\Delta s(x_B) ~ \Delta D(z)} {s(x_B)  ~ D (z)}.
 \label{res}
\eeq

At present,  very little is known for the strange quark
fragmentation functions $D (z)$ and $\Delta  D(z)$, so we
are unable to make quantitative predictions on the
longitudinal polarization of the semi-inclusive $\Lambda$
particles.  However, $D (z)$ and $\Delta  D(z)$ can be measured
independently  from the inclusive $\Lambda$ production
by  electron-positron annihilation near the $Z$ resonance,
as Burkardt and Jaffe  \cite{BJ} have suggested. Once
such an experiment is completed,  the obtained  data on
$D (z)$ and $\Delta  D(z)$  can be evolved by the Altarelli-Parisi
equations \cite{AP} onto the energy
scale we are interested in. Then, with $D (z)$ and $\Delta  D(z)$
as inputs,  the measurements of the polarization of the
semi-inclusive $\Lambda$'s  will supply us with data
on $\Delta s(z)$.

Here we  should note that the  sign of the $\Lambda$ polarization,
let alone its precise value, will greatly  help us gain insight
into the strange content of the proton. From Eq. (\ref{res}),
we know that the  sign of the $\Lambda$ polarization is
determined by  the product of $\Delta s(x_B)$ and $\Delta D(z)$.
Naively, one may anticipate that  the  probability for the
fragmented $\Lambda$ hyperon to have the same helicity as its
parent strange quark  is over that to possess the opposite helicity,
since  the particles in the current fragmentation region are
almost collinear with the parent quark.  If this is the case,
and if the strange quark in the proton is
really polarized as the EMC experiments indicated,  the
$\Lambda$  polarization will also be  negative.  However,
such a conclusion is unsafe since  our experience  with polarized
deeply inelastic  structure functions suggest that $\Delta D(z)$
might be negative as well. If so, the polarization of the
semi-inclusive $\Lambda$ hyperons will be positively polarized.
Anyway, if  the $\Lambda$ polarization is measured to
be negative, it will increase our confidence that
the strange  content of the proton is  negatively polarized.
It is based on this fact that
we suggest  measuring the $\Lambda$  polarization in
the semi-inclusive production  process, with the lepton beam unpolarized
and the proton target longitudinally polarized.  Such an
experiment is in principle feasible, since people have accumulated
relatively rich experience in  measuring the $\Lambda$ polarization
\cite{reports,prd}.

Before we end up this Letter,  let us note that in  reaching
our conclusion,  we have suppressed some
secondarily important ingredients for simplicity. In general,
they can classified into the radiative corrections in the hard
partonic scattering  subprocesses and the higher twist effects
due to the intrinsic  transverse   quark momentum  in the proton
and the parton correlations in the course of the strange quark
fragmentation. Integrating them into the scheme is straightforward but
tedious,  which should deserve a long publication. We anticipate
that radiative corrections and higher twist effects can
cause quantitatively significant  modifications to our conclusion.
At any rate, the prediction for the sign of the $\Lambda$ polarization,
based on the interpretation of the Ellis-Jaffe sum rule breaking,
will be unchanged in the more complete factorization approach.

In summary,  we  have examined the implication of the
possible large negative polarization of strange quarks in the proton
to the $\Lambda$ polarization in the
semi-inclusive deep inelastic lepton proton scattering. A
qualitative result is that if the strange sea in the
proton is negatively polarized and the $s\to \Lambda$ fragmentation
can transfer the  strange quark helicity onto the fragmented
$\Lambda$, the semi-inclusive $\Lambda$
will be negatively polarized as well. Considering that  the
strange sea content in the proton is closely related to the
proton spin puzzle raised by the Ellis-Jaffe sum rule violation,
we suggest measuring the polarization of the semi-inclusive $\Lambda$'s,
which  will be feasible in the next generation of experiments, for
example,  HERMES at DESY and HELP at CERN.  The proposed
experiment is anticipated to shed some light
on our understanding  of the proton structure.

{\it Note added}: After the submission of this paper,
we learned of  a manuscript  by J. Ellis, D. Kharzeev, and
A. Kotzinian (hep-ph/9506280), which contains a
very similar idea in this paper.

\end{document}